\documentclass[%
reprint,
superscriptaddress,
showpacs,preprintnumbers,
 amsmath,amssymb,
 aps,
pra,
]{revtex4-1}

\usepackage{graphicx}   
\usepackage{dcolumn}
\usepackage{bm}
\usepackage{hyperref}
\graphicspath{{fig/}}   

\usepackage{epstopdf}

\usepackage{braket}
\usepackage{siunitx}
\usepackage{changes}

\binoppenalty=\maxdimen
\relpenalty=\maxdimen
\mathchardef\mhyphen="2D

\usepackage[plain]{fancyref}

\newcommand{\ie}{i.e.\@ }
\newcommand{\CO}{(Color Online) }

\begin{document}
\title{Transmon Qubit in a Magnetic Field: \\Evolution of Coherence and Transition Frequency}

\author{Andre Schneider}	
\email{andre.schneider@kit.edu}
\author{Tim Wolz}	
\author{Marco Pfirrmann}	
\author{Martin Spiecker}
\author{Hannes Rotzinger}
	\affiliation{Institute of Physics, Karlsruhe Institute of Technology, 76131 Karlsruhe, Germany}
\author{Alexey V. Ustinov}	
	\affiliation{Institute of Physics, Karlsruhe Institute of Technology, 76131 Karlsruhe, Germany}
	\affiliation{Russian Quantum Center, National University of Science and Technology MISIS, 119049 Moscow, Russia}
\author{Martin Weides}		
	\email{martin.weides@glasgow.ac.uk}
	\affiliation{Institute of Physics, Karlsruhe Institute of Technology, 76131 Karlsruhe, Germany}
	\affiliation{James Watt School of Engineering, University of Glasgow, Glasgow G12 8LT, United Kingdom}

\date{\today}


\begin{abstract}
We report on spectroscopic and time-domain measurements on a fixed-frequency concentric transmon qubit in an applied in-plane magnetic field to explore its limits of magnetic field compatibility.
We demonstrate quantum coherence of the qubit up to field values of $B=\SI{40}{\milli\tesla}$, even without an optimized chip design or material combination of the qubit.
The dephasing rate $\Gamma_\varphi$ is shown to be not affected by the magnetic field in a broad range of the qubit transition frequency.
For the evolution of the qubit transition frequency, we find the unintended second junction created in the shadow angle evaporation process to be non-negligible and deduce an analytic formula for the field-dependent qubit energies.
We discuss the relevant field-dependent loss channels, which can not be distinguished by our measurements, inviting further theoretical and experimental investigation.
Using well-known and well-studied standard components of the superconducting quantum architecture, we are able to reach a field regime relevant for quantum sensing and hybrid applications of magnetic spins and spin systems.
\end{abstract}


\maketitle

\section{\label{sec:introduction}{Introduction}}
Superconducting quantum circuits render a versatile platform for realizing circuit quantum electrodynamic (cQED) systems.
Such systems are used in various applications as they offer a flexible and engineerable toolset to build a physical model system and employ it to study quantum mechanics in depth. 
They can also be used for interaction and characterization of other quantum systems and turn out to be a useful tool for investigation.
Superconducting quantum bits are a promising candidate for quantum computing and quantum simulation \cite{Popkin2016}, as well as for the emerging field of quantum sensing \cite{Degen2017}, which becomes more and more important with the fast growing number of quantum systems that are subject to current research. 
Superconducting qubits are here used to study the characteristics and dynamics of unknown systems in the quantum regime and are therefore a valuable sensing tool.

Applications like quantum sensing of magnetic excitations \cite{Degen2017}, creating and harnessing Majorana Fermions \cite{Mourik2012} or 
quantum cavity magnonics \cite{Tabuchi2015,Pfirrmann2019}
expose the qubits to magnetic fields.
In particular superconducting qubits are intrinsically vulnerable to magnetic fields.
So far, in the literature only influences of small magnetic fields on the order of \SI{100}{\micro\tesla} have been studied, where even a slight improvement of their coherent behavior for very small fields could be found due to the creation of quasiparticle traps by entering flux vortices \cite{Wang2014}.
However, the general consent is to screen magnetic fields as best as possible and a multi-layered shielding based on permalloy and superconductors 
is commonly used
\cite{Kreikebaum2016,Flanigan2016}. 
To our knowledge, no published effort has been spent to study the limits of magnetic field compatibility of standard Josephson junction (JJ) qubits. 
In fact, they have been assumed to break down at very little fields and other, more stable junctions, such as the proximitized semiconducting nanowire, have been introduced to circumvent this limitation \cite{Luthi2018}.

In this article however, we study the magnetic field properties of a conventional Josephson tunneling barrier junction qubit for in-plane magnetic fields up to \SI{40}{\milli\tesla}, which is well above the saturation field for magnets like permalloy, opening opportunities in hybrids of quantum circuits and magnetic materials.

This letter starts with the investigation of the magnetic field dependence of the qubit's transition frequencies, where we find an analytic formula.
In the following, we study the coherence time under the influence of a magnetic field and discuss different field-dependent loss channels.
This behavior is reproducible and symmetric with respect to the applied fields up to $B=\pm\SI{20}{\milli\tesla}$. Going to stronger fields, we demonstrate measurable coherence times up to $B=\SI{40}{\milli\tesla}$,
and remanently suppressed coherence when decreasing the field again.
In the last section we analyze the pure dephasing rate, which we find to be independent from the magnetic field.

\section{Sample and Setup}
The qubit used for this experiment is a single-junction concentric transmon \cite{BraumuellerAPL16}, which was already described in Ref.\@ \cite{Schneider2018}.
Its capacitance pads are made from low-loss TiN and the junction is an Al/$\mathrm{AlO_x}$/Al structure, fabricated by shadow angle evaporation.
The sample is placed in a copper box and mounted to the base stage of a dilution refrigerator at a temperature of about $\SI{30}{\milli\kelvin}$. 
It reaches into a solenoid fixated at the still stage. 
The sample was aligned to the solenoid for an in-plane orientation of the field by eye, leaving a probability for small out-of-plane field components at the sample. 
Due to the structured superconductor on the chip and the resulting flux-focusing \cite{BuckelKleiner} leading to an inhomogeneous magnetic field, we assume that an ideal in-plane configuration over the whole chip is hard to achieve.
This especially holds true when looking at future applications, where a possible local magnetic environment produces stray fields.
 
For the measurements, a time-domain as well as a spectroscopy setup is used, which are described in Appendix \ref{ap:setup} together with the cryogenic setup. 
Data acquisition and analysis are performed via \textsc{Qkit} \cite{qkit}.

To infer the qubit state, we observe the dispersive frequency shift of 
a $\lambda/2$ resonator coupled to the qubit by $g/2\pi = \SI{71.5}{\mega\hertz}$.
The microstrip resonator is made from low-loss TiN,
with initial frequency $\omega_\mathrm{r,0}/2\pi = \SI{8.573}{\giga\hertz}$ and 
internal quality factor around $Q_\mathrm{i,0}=\SI[separate-uncertainty]{5100\pm120}{}$, extracted by a circle fit \cite{Probst2015}.
When changing the magnetic field, we see a reproducible field-dependent change of  $Q_\mathrm{i}$ and $\omega_\mathrm{r}$ which is hysteretic due to the creation and annihilation of flux vortices in the material \cite{Bothner2012}, see Appendix \ref{ap:res} for more data.
The reducing quality factor involves a decreased signal to noise ratio (SNR) for our measurements, making it harder to find the qubit transition frequency.

\section{Qubit transition frequency in an in-plane magnetic field}
\begin{figure}[tb]
	\includegraphics[width=\columnwidth]{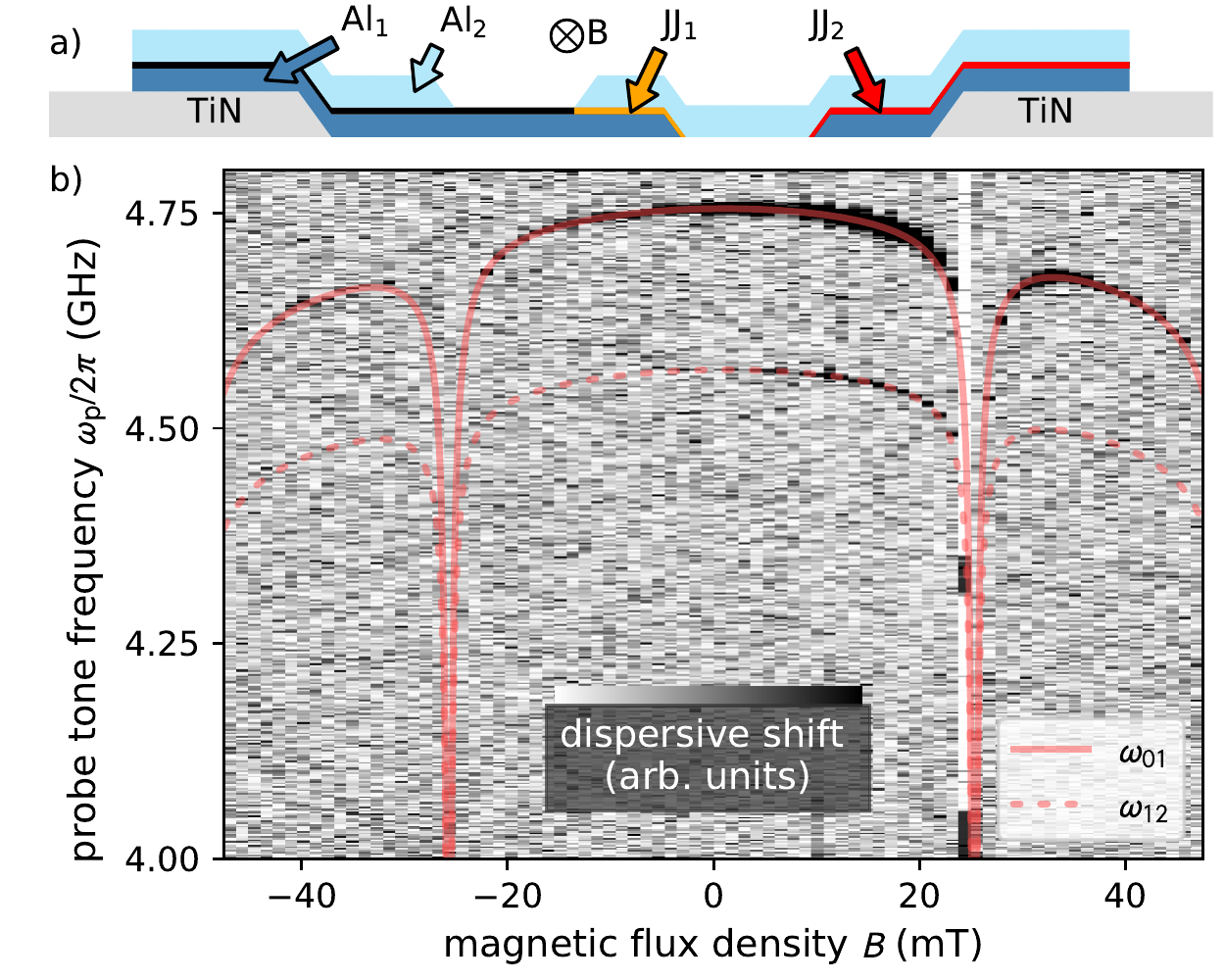}
	\caption{
		\CO (a) Sketch of the shadow evaporated junction on top of the TiN capacitance pads. The shaded areas are the two aluminum layers with an $\mathrm{AlO_x}$ barrier in between, forming two JJs (not to scale). $\mathrm{JJ_2}$ is the spurious junction created by the process.
		(b) Qubit transition frequency vs applied magnetic field. A clear periodic behavior is obvious but can not be explained by a single JJ. The colored lines show the qubit transition frequencies calculated from \fref{eq:H}. $\mathrm{JJ_1}$ gives the overall envelope and $\mathrm{JJ_2}$ the periodic minima.
	}
	\label{fig:two_jj}
\end{figure}
In the following, we study the qubit transition frequency under the influence of an applied in-plane magnetic field. 
It is known that this field suppresses the critical current $I_\mathrm{c}$ of a Josephson junction periodically, where the shape follows a Fraunhofer pattern \cite{Barone}.
In contrast to this expectation with $\omega_{01} \propto \sqrt{I_\mathrm{c}}$, our measurement data show a flat top at $\Phi =0$, a much steeper slope at $\Phi = \Phi_0$, and an overall envelope, i.e. the first side maxima are not as high as the main maximum (see \fref{fig:two_jj}b)).

\subsection{JJ Fabrication Scheme}
Due to the JJ fabrication by shadow angle evaporation, two tunnel junctions exist in series, see \fref{fig:two_jj}a).
The current flows from the TiN layer through the lower Al layer, via the designed tunnel junction ($\mathrm{JJ_1}$) to the top layer, and then passes the oxide barrier ($\mathrm{JJ_2}$) to reach the second TiN electrode. 
$\mathrm{JJ_2}$ is much larger in area, has a very high critical current $I_{\mathrm{c},2} \gg I_{\mathrm{c},1}$, and is therefore commonly neglected for the qubit properties.
Its large cross section however gives an increased sensitivity to the applied magnetic flux.
This second junction can be avoided by a shunting bandage \cite{Dunsworth2017} or overlap junctions \cite{Wu2017}. 
The third JJ on the left side of \fref{fig:two_jj}a) however would only shunt the $\mathrm{Al_1}$ layer and can therefore be neglected as long as the inductance of the lower layer is negligible.

\subsection{Qubit Hamiltonian with Two JJ in Series}
To calculate the current-phase-relation for two JJ in series, we start with two junctions in series, having critical currents $I_{\mathrm{c},1},\, I_{\mathrm{c},2}$ and phases $\varphi_1,\, \varphi_2$. 
It is clear that the current through them is the same and the total phase adds up:
\begin{eqnarray}
&& \varphi_\Sigma = \varphi_1 + \varphi_2\nonumber\\
I( \varphi_\Sigma) &=& I_{\mathrm{c},1} \sin \varphi_1 = I_{\mathrm{c},2} \sin \varphi_2.\nonumber
\end{eqnarray}
We introduce the ratio $ r  = I_{\mathrm{c},2}/ I_{\mathrm{c},1}$ between the junctions' critical currents and assume $r \geq 1$ without loss of generality. From this we follow
\begin{eqnarray}
\tan \varphi_2 &=&  \frac{\sin\varphi_\Sigma}{r+\cos\varphi_\Sigma}.\nonumber
\end{eqnarray}
The overall current-phase-relation is thus given by
\begin{eqnarray}
\label{eq:curphase}
	I (\varphi_\Sigma)
	=  I_{\mathrm{c},2}\sin \left(  
	\arctan\frac{\sin\varphi_\Sigma}{r+\cos\varphi_\Sigma}  \right),
\end{eqnarray}
resulting in the system Hamiltonian
\begin{eqnarray}
H = 4 E_\mathrm{C}N^2  - E_\mathrm{J,1}    \sqrt{r^2+2r\cos\varphi+1}, \nonumber
\end{eqnarray}
where the exact derivation can be found in Appendix \ref{ap:twojj}. For the approximate transmon Hamiltonian we get
\begin{eqnarray}
\label{eq:H}
H &\approx& \sqrt{8E_\mathrm{C} E_\mathrm{J,1}\frac{r}{r+1}}a^\dag a - \frac{r^2-r+1}{(r+1)^2}\,\frac{E_\mathrm{C}}{2}a^\dag a(a^\dag a+1). \qquad
\end{eqnarray}
Here, $a^\dag$ and $a$ are the harmonic oscillator creation and annihilation operators, $E_\mathrm{C}=\mathrm{e}^2/2C = \SI{190}{\mega\hertz}$ is the charging energy, and $E_\mathrm{J} = I_\mathrm{c} \Phi_0 /2\pi$ is the Josephson energy.
For the limit of $r \rightarrow \infty$, where $\mathrm{JJ}_1$ is dominating (i.e.\@ limiting) the circuit, this formula goes back to the unperturbed approximated transmon Hamiltonian \cite{Koch_TransmonPRA07}.
We emphasize that the transmon's anharmonicity decreases if the two junctions are comparable in $I_\mathrm{c}$, i.e.\@ for $r=1$ the maximum anharmonicity is reduced by a factor of 4.

With \fref{eq:H} we calculate the transmon spectrum and find good agreement with the measured data in \fref{fig:two_jj}b).
For the individual junctions,  $I_\mathrm{c} = I_\mathrm{c}^0 \left|  \mathrm{sinc} \frac{B-B_\Delta}{B_{\Phi_0}}  \right|$ is assumed \cite{Barone}, where $B$ is the applied magnetic field, $B_\Delta$ is a constant offset due to background fields, and $B_{\Phi_0}$ is a measure for the field periodicity of the corresponding junction, see \fref{tab:qparams}.

The flux penetrating the JJ is given by $\Phi = BA =\allowbreak B (d+2\lambda_\mathrm{L})l$ with $A$ the effective junction cross section area, $d = \SI{1}{\nano\meter}$ the thickness of the oxide barrier, $\lambda_\mathrm{L} = \SI{16}{\nano\meter}$ the London penetration depth of Al, and $l$ the length of the junction. From that we can calculate the junctions' lengths as $l_1 = \SI{209}{\nano\metre}$  and $l_2 = \SI{2.46}{\micro\meter}$, agreeing well with the design parameters. The reduction of the superconducting gap additionally creates an envelope to the curve, which is discussed in Appendix \ref{ap:gap}.
The existence of $\mathrm{JJ}_2 $ implies that the insulating barrier exists consistently over the large junction area and therefore
demonstrates the good quality of the oxide film.
\begin{table}
\caption{Parameters of the two Josephson junctions. $B_\Delta$ is the offset field, $B_{\Phi_0}$ the periodicity, and $l$ the length of the junction.}
\label{tab:qparams}
\begin{tabular*}{\columnwidth}{c@{\extracolsep{\fill}}cccc}
	\hline\hline
 	& $E_\mathrm{J}/h\ (\si{\giga\hertz})$ & $B_\Delta\ (\si{\milli\tesla})$ & $B_{\Phi_0}\ (\si{\milli\tesla})$ & $l\ (\si{\nano\metre})$ \\ \hline 
	$\mathrm{JJ}_1$ & 16.15 & 1.8 & 300 & 209\\
	$\mathrm{JJ}_2 $ & 300 & -0.2 & 25.5 & 2460 \\
		\hline\hline
\end{tabular*} 

\end{table}

\section{Qubit Coherence Times}
\subsection{Measurement Sequence}
To measure the coherence times of the transmon qubit in a magnetic field, we construct a measurement sequence that ramps the field to a specified value, scans the readout tone to find the shifted resonator frequency $\omega_\mathrm{r}(B)$, and scans the probe tone to find the qubit transition $\omega_{01}$. A Rabi sequence to find the length $t_\pi$ of a $\pi$-pulse is applied to the qubit and finally a sequence of $T_1, T_2^\mathrm{Ramsey}$ and/or $T_2^\mathrm{Echo}$ measurements is executed to get the desired measurement values. 
The number of averages and points per trace is reduced to perform the whole sequence for one field value within 10 minutes, despite the low SNR.

\begin{figure}[tb]
	\includegraphics[width=\columnwidth]{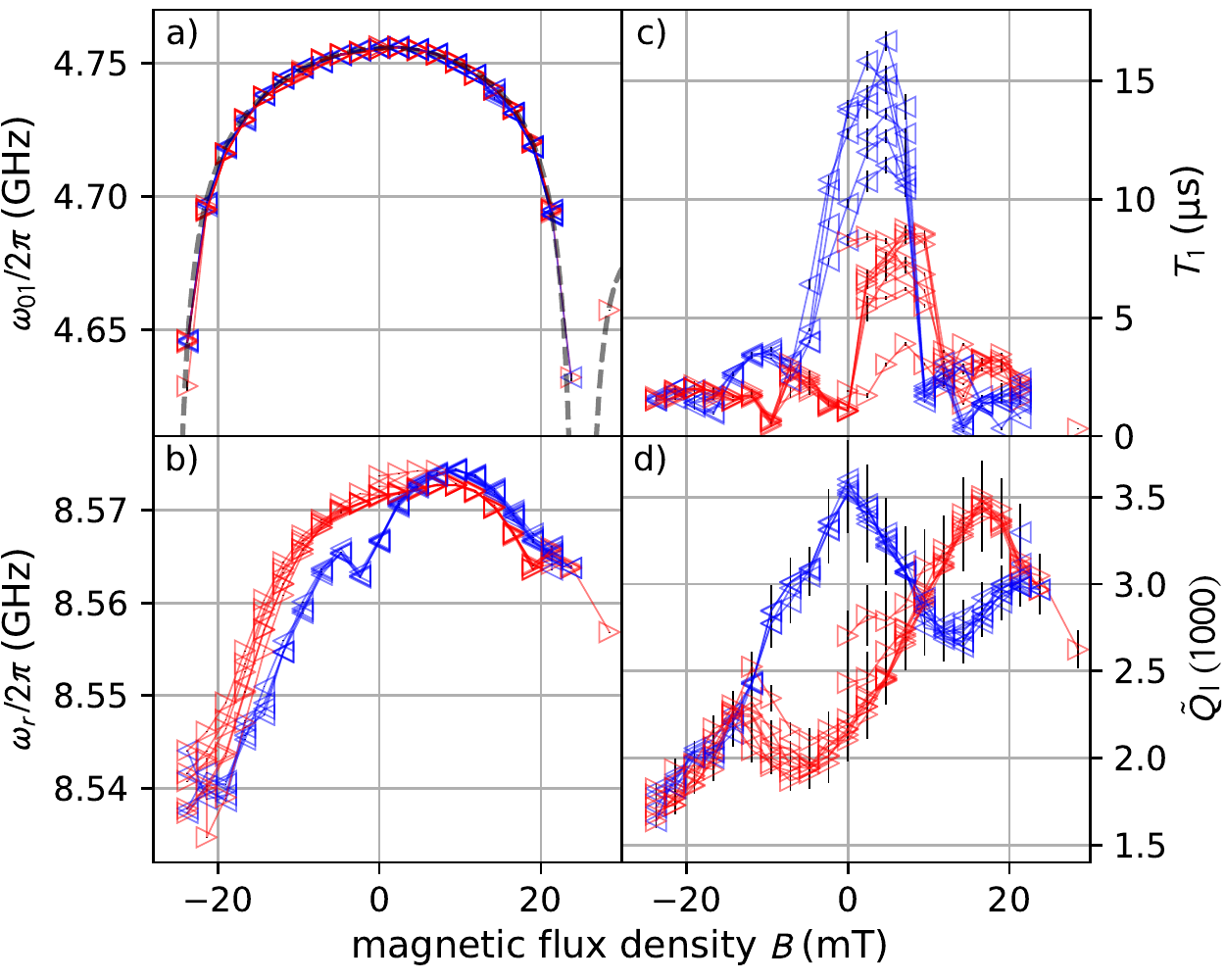}
	\caption{
		\CO
		Time domain measurements: Red (blue) triangles pointing to the right (left) are data points taken on an up (down)sweep.
		The measured qubit transition frequency (a) shows no hysteresis and corresponds well with our theoretical description (dashed line). 
		A clear hysteresis can be seen for the resonator frequency (b), the qubit $T_1$ time (c) and the resonator quality factor (d). Due to the nonlinear amplitude detection in the time domain setup, the displayed quality factor $\tilde{Q}_\mathrm{l}$ is only an indicative value for $Q_\mathrm{l}$ and properly measured values can be found in Appendix \ref{ap:res}.
	}
	\label{fig:coherence_quad}
\end{figure}

The results of multiple sweeps in the range of $B = \pm\SI{23.7}{\milli\tesla}$ are shown in \fref{fig:coherence_quad} where red (blue) triangles mark the points taken on an up (down) sweep of the magnetic field.
The qubit transition frequency $\omega_{01}$ follows \fref{eq:H} and shows no hysteresis.

\subsection{Loss Mechanisms}
The $T_1$ time of the qubit shows a pronounced maximum at low fields and is clearly different on up and down sweeps. 
To characterize this behavior, we separate the losses of the system as
\begin{eqnarray}
\frac{1}{T_1} = \Gamma_1 = \Gamma_\mathrm{hyst} + \Gamma_\mathrm{non\mhyphen hyst} + \Gamma_\mathrm{const},\nonumber
\end{eqnarray}
where $ \Gamma_\mathrm{hyst} $ accounts for loss mechanisms showing a hysteretic field dependence, $ \Gamma_\mathrm{non\mhyphen hyst} $ collects losses that depend directly on the magnetic field strength, and the losses associated with $ \Gamma_\mathrm{const} $ do not depend on the magnetic field.

We attribute the hysteretic loss mechanisms $ \Gamma_\mathrm{hyst} $ mainly to the dissipation introduced by the entering of flux vortices in the thin film superconductor and their movement due to the oscillating RF current, which was already observed for superconducting resonators \cite{Bothner2012}.
The quality factor of a resonator is a measure for its excitation lifetime and is therefore equivalent to the $T_1$ time for the qubit.
The shapes and signs of the envelopes of $\tilde{Q}_\mathrm{l}$ and $T_1$ are generally similar (\fref{fig:coherence_quad}c) and d)), as the two mainly consist of the same material.
The observed mismatch can be attributed to their very different geometries and current distributions.
From the large aspect ratio of the qubit island with $\SI{554}{\micro\metre}$ diameter and $\SI{40}{\nano\metre}$ thickness we conclude that the vortices are mainly generated perpendicular to the film.

Non-hysteretic losses $ \Gamma_\mathrm{non\mhyphen hyst} $ are mainly attributed to the dissipation through excitations of the superconductor, i.e.\@ quasiparticles (QP). A linear relation between the QP density and  $ \Gamma_\mathrm{QP} $ has been demonstrated \cite{Wang2014} as well as 
a quadratic dependence of the QP density on the magnetic field \cite{Kwon2018} and a reduced QP recombination rate in magnetic fields \cite{Xi2013}.
The QP density is not reported to have a hysteretic dependence on the effective field and the relaxation to an equilibrium QP density is expected to happen within a few $\si{\micro\second}$. The hysteretic vortex configuration however affects the effective field in the superconductor and therefore the QP density.

A small number of pinned flux vortices can also decrease the number of QPs,
as the normal conducting cores of the vortices act as QP traps.
This can be seen in an increasing $T_1$ time for $B -B_\mathrm{offs}>0$ ($<0$) on the up (down) sweeps.
We attribute the average offset field $B_\mathrm{offs} = \SI{8.5}{\milli\tesla}$ to the presence of stray fields from magnetized components around the qubit chip, 
which are partially aligned perpendicular to the chip. 
Taking into account a small misalignment between coil field and chip of about $\alpha \approx 3 \si{\degree}$, an applied field of $\SI{8.5}{\milli\tesla}$ would compensate a perpendicular magnetic field of  $B \approx 450\,\si{\micro\tesla}$, being on the order of typical stray fields.
Measurements after a cycle of the sample temperature above $\SI{7}{\kelvin}$ showed $T_1$ times on the order of few $\si{\micro\second}$, being comparable to the values for zero applied field at upsweeps and demonstrating the constant background field.

Relaxation sources like Purcell loss, radiative losses, and losses to two-level-systems in the junction and on the surface of the qubit islands do not depend on the magnetic field and are represented by $ \Gamma_\mathrm{const} $.

\begin{figure}[tb]
	\includegraphics[width=\columnwidth]{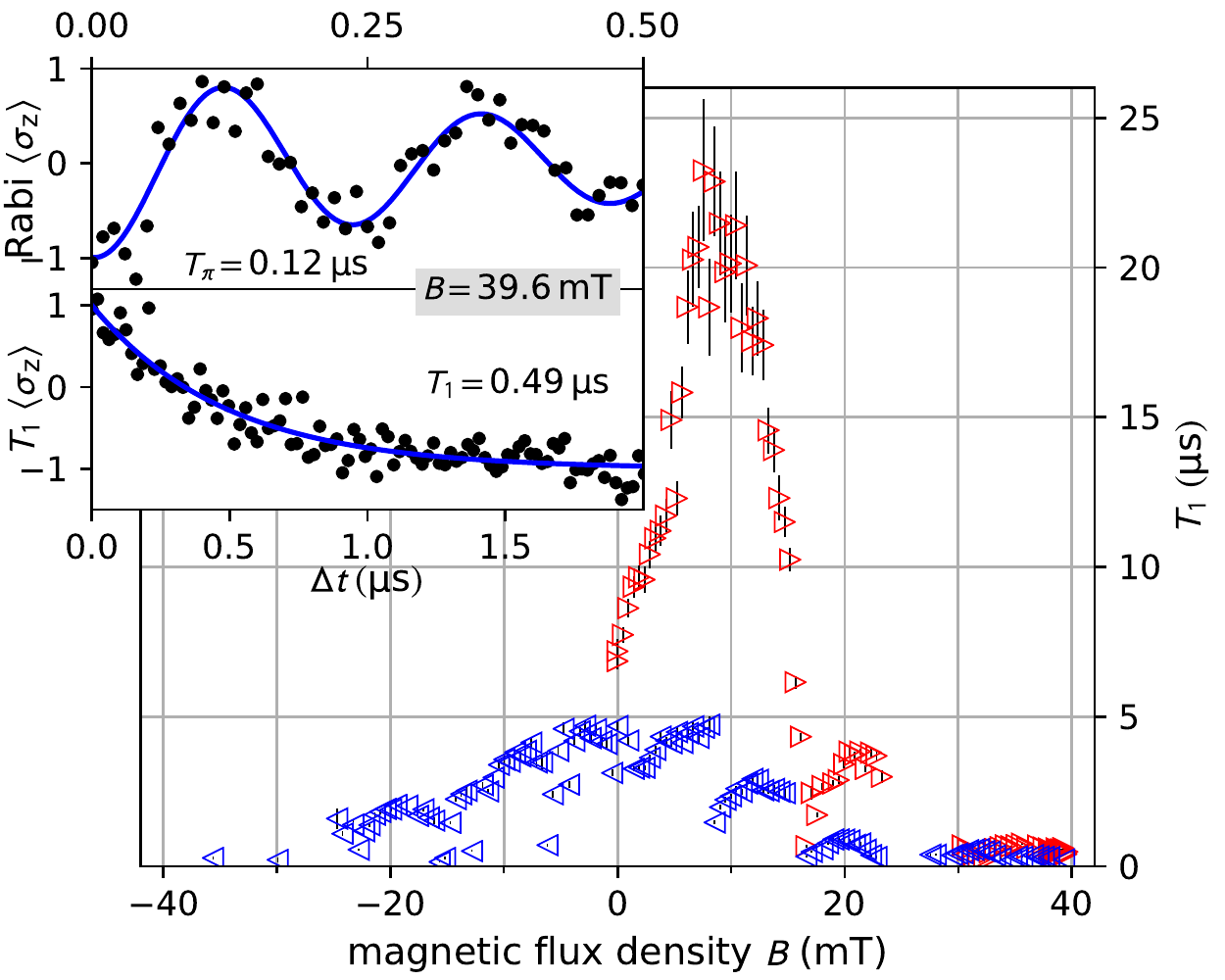}
	\caption{
		\CO
		Qubit coherence up to $\SI{40}{\milli\tesla}$. A pronounced maximum is visible for low fields in the upsweep. In the down sweep, no maximum is observed but the $T_1$ data show discrete jumps at different field values, attributed to the movement or annihilation of a flux vortex. The inset shows measured data for a Rabi and $T_1$ measurement at the highest $B$ value. The shown values for $\langle {\sigma_z} \rangle$ are nomalized to the accessible qubit values, \ie an increased residual population is calibrated away.
	}
	\label{fig:highfield}
\end{figure}

\subsection{Increased Magnetic Field}
While the previous measurements have been hysteretic but repeatable, we now further increase the magnetic field to
 stronger fields and  see quantum coherence of the qubit up to values of $B_\mathrm{appl} \approx \SI{40}{\milli\tesla}$ (\fref{fig:highfield}).
Although $T_1 = \SI{0.49}{\micro\second}$ is significantly reduced, we can observe Rabi oscillations and an exponential decay after a $\pi$ pulse, as demonstrated in the inset of \fref{fig:highfield}.
At these fields, the quality factor $Q_\mathrm{l}$ of the resonator is significantly reduced, explaining the low SNR.
Together with the decreased $T_1$, $T_2$ times and the resulting broadening of the qubit linewidth, the qubit transition could not be tracked for even higher magnetic fields, as visible in \fref{fig:two_jj}b).

The subsequent down sweep does not show a pronounced maximum as before but only a slight increase in $T_1$ over a broader range. We also see a fine structure in the data, showing multiple drops in $T_1$ which coincide with the onset of a deviation from the Fraunhofer pattern, followed by a jump in frequency.
We attribute this effect to the presence of flux vortices in the qubit islands due to the previously applied high fields.
Their local field influences the field seen by the junction and therefore qubit frequency and coherence.

\section{Pure dephasing rate}
\begin{figure}[tb]
	\includegraphics[width=\columnwidth]{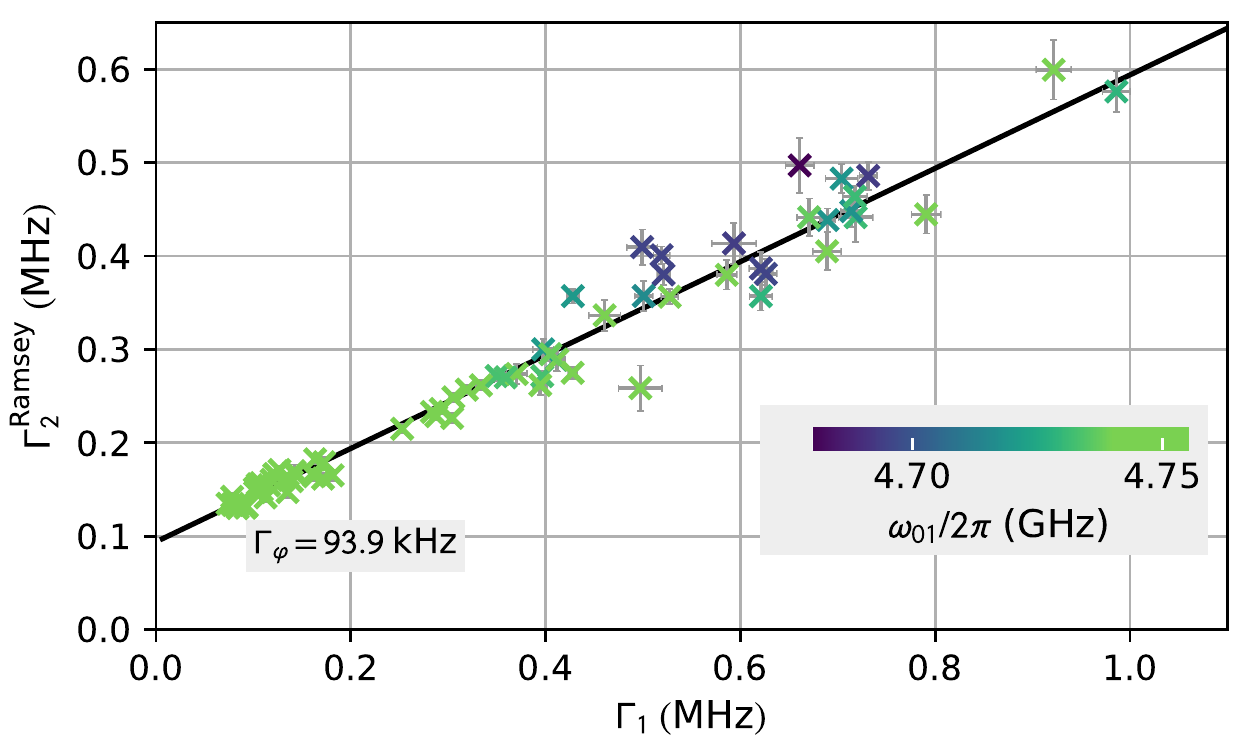}
	\caption{
		\CO
		Pure dephasing rate and correlation between $\Gamma_2^\mathrm{Ramsey}$ and $\Gamma_1$. 
		The solid line depicts the expected relation for a constant dephasing, invariant of the magnetic field, where the pure dephasing rate $
		\Gamma_\varphi = \SI{93.9}{\kilo\hertz}$ is extracted from the data.
		The color of the data points corresponds to the qubit transition frequency and therefore its sensitivity to flux noise.
	}
	\label{fig:tphi}
\end{figure}
To calculate the pure dephasing rate of the qubit $\Gamma_\varphi$  from measured values, we take
\begin{eqnarray}
\label{eq:dephasing}
\Gamma_2^\mathrm{Ramsey} = \frac{1}{2} \Gamma_1 + \Gamma_\varphi,
\end{eqnarray}
where $\Gamma_1 = 1/T_1$ and $\Gamma_2^\mathrm{Ramsey} = 1/T_2^\mathrm{Ramsey}$ are the decay and Ramsey dephasing rates.
In order to have physically connected $\Gamma_1 $ and $\Gamma_2^\mathrm{Ramsey}$ rates, we acquire the measurement points for both rates in turn, so that temporal fluctuations of the qubit properties influence both measurements likewise \cite{Schloer2019}.
The resulting data are shown in \fref{fig:tphi} and fits to a straight line of a constant pure dephasing $\Gamma_\varphi = \SI{93.9}{\kilo\hertz}$.
For the regions of a steep slope of $\omega_{01}(B)$, a higher dephasing rate would be expected due to the stronger sensitivity to flux noise. 
However, a clear correlation between $\Gamma_\varphi$ and $\omega_{01}$ can not be seen from the data.
The causality between noise in the solenoid current $S_I$ and the resultant $\Gamma_\varphi^{I}$ is given by \cite{Ithier2005}: 
\begin{eqnarray}
\Gamma_\varphi^{I}=\pi \left(  \frac{\partial \omega_{01}}{\partial I}  \right)_z^2 S_I (\omega \ll k_\mathrm{B}T),
\end{eqnarray}
where the relevant scale for $\omega$ is the time between the Ramsey pulses, being on the order of $\omega/2\pi \approx \SI{100}{\kilo\hertz}$.

From \fref{eq:H} we calculate  a slope of the qubit transition frequency of 
$\left(  {\partial \omega_{01}}/{\partial I}  \right)/ 2\pi  =\SI{652}{\mega\hertz\per\ampere}$
at
$B=\SI{21}{\milli\tesla}$
and 
$\omega_{01}(B)/2\pi = \SI{4.70}{\giga\hertz}$.
Considering the measured power spectral density of our current source  $S_I \approx\SI{E-15}{\ampere\squared\per\hertz}$,
this results in $\Gamma_\varphi^I = \SI{53}{\kilo\hertz}$, well below our measured $\Gamma_\varphi$. We conclude that for the main part of the qubit spectrum,  $\Gamma_\varphi$ is not limited by current fluctuations or other fluctuating stray magnetic fields and
the qubit coherence is governed by a magnetic field independent dephasing rate.

\section{Conclusion}
In this article, we demonstrated the quantum coherence of a superconducting transmon qubit in magnetic fields up to a flux density of \SI{40}{\milli\tesla}, which increases their usability range as versatile sensors and is a promising finding for future developments of superconductor-magnet-hybrid devices.
The influence of the second, spurious junction in circuits fabricated by shadow angle evaporation was shown, where its large area gives rise to a higher sensitivity to in-plane magnetic fields.
To calculate the influence of this additional junction on the qubit transition frequency, an analytic formula for the approximated transmon Hamiltonian featuring two serial junctions was derived.
Finally we studied the pure dephasing rate and found it to be mainly independent of the magnetic field.

\begin{acknowledgments}
The authors are grateful for quantum circuits provided by D. Pappas, M. Sandberg, and M. Vissers. 
This work was supported by the European Research Council (ERC) under the Grant Agreement 648011, Deutsche Forschungsgemeinschaft (DFG) within projects WE4359/7-1 and INST 121384/138-1, and the Initiative and Networking Fund of the Helmholtz Association.
We acknowledge financial support by the Carl-Zeiss-Foundation (A.S.) and the Helmholtz International Research School for Teratronics (T.W. and M.P.).
A.V.U. acknowledges partial support from the Ministry of Education and Science of the Russian Federation in the framework of the contract No.\@ K2-2016-063.
\end{acknowledgments}

\appendix
\clearpage

\renewcommand{\thefigure}{S\arabic{figure}}
\renewcommand{\theequation}{S\arabic{equation}}
\setcounter{figure}{0}
\setcounter{equation}{0}

\section*{Supplementary Material for ``Transmon Qubit in a Magnetic Field: Evolution of Coherence and Transition Frequency''}
\section{Resonator in a Magnetic Field}
\label{ap:res}
\begin{figure}[tb]
	\includegraphics[width=\columnwidth]{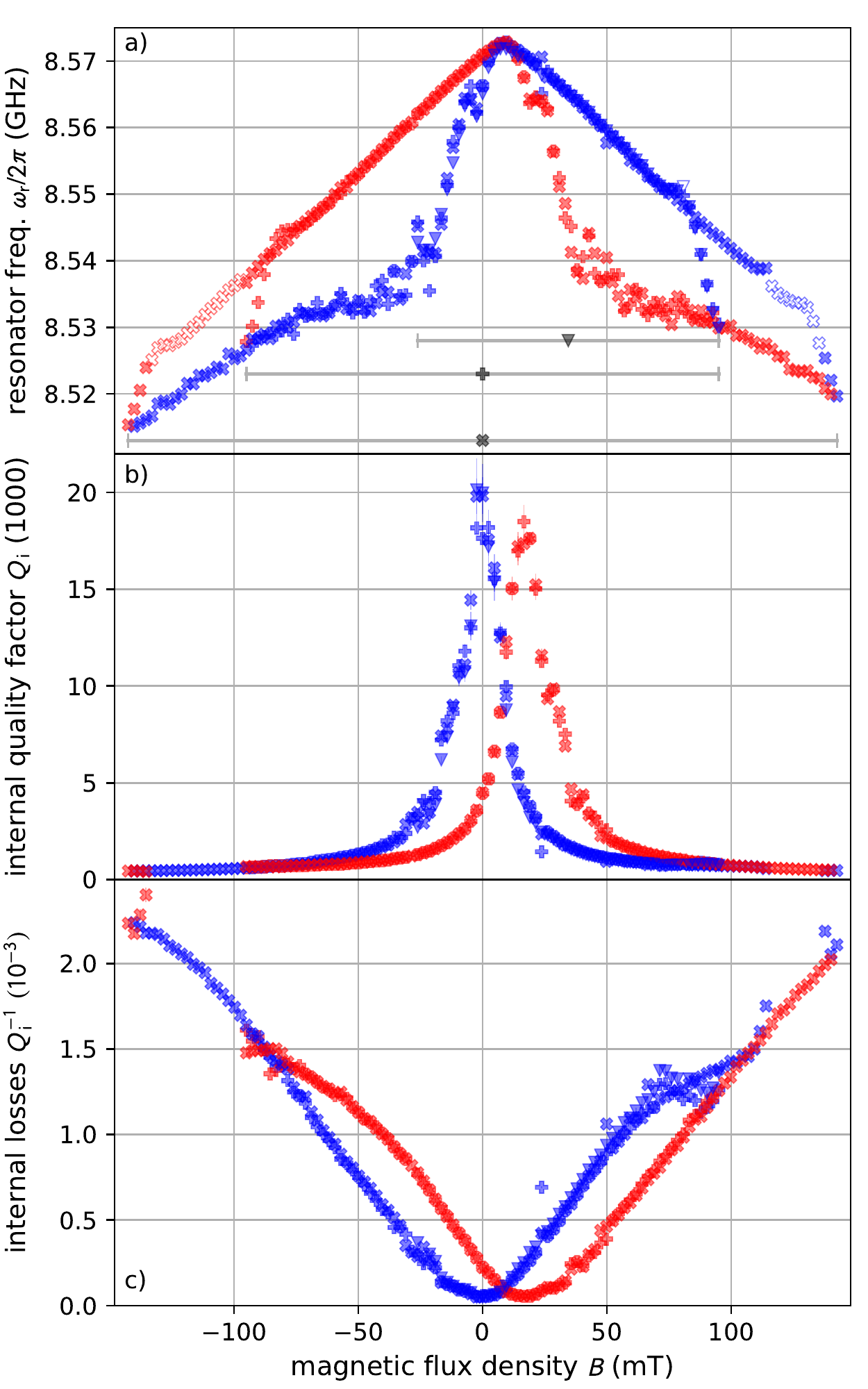}
	\caption{
		Resonance frequency (a) and internal quality factor (b) of the resonator under the influence of an in-plane magnetic field. 
		Red (blue) points represent data taken on an up (down) sweep of the magnetic field. 
		Up and down sweeps with the same symbol are taken in series, where the down sweep was first.
		The values for the closed symbols are extracted by a circle fit \cite{Probst2015}. A simple Lorentzian fit to the measured amplitudes was used for the open symbols due to the weak phase signal.
		The gray bars in a) mark the sweep range for the magnetic field with the corresponding symbols.
		Plot (c) of $\Gamma_\mathrm{i} = Q_\mathrm{i}^{-1}$ is for comparison with published data \cite{Bothner2012}.
	}
	\label{fig:res}
\end{figure}

The measured data on the field dependence of resonance frequency and quality factor of our readout resonator correspond well to already published data \cite{Bothner2012} for an in-plane magnetic field. 
In their publication, the loss rate $\Gamma_\mathrm{i}$ is calculated by using the classical Bean model \cite{Bean1962,Bean1964} and their simulation matches our data very well for the case of a weakly inhomogeneous RF current distribution.
Although \fref{fig:res}c) suggests that the resonator is completely interspersed with flux vortices at $B \approx \SI{100}{\milli\tesla}$, a closer look in the data does not support this statement, as the phase signal becomes very weak in the region of $\left| B \right| > \SI{100}{\milli\tesla}$ and the circle fit \cite{Probst2015} does not converge. 
Fitting the measured data with a Lorentzian still shows a difference between up and down sweep for the loaded quality factor (data not shown).

From the circle fit data, we extract the coupling quality factor to be $Q_\mathrm{c}=\SI[separate-uncertainty]{9.3\pm 1.3}{\mega\hertz}$. 
This quantity is defined by the geometric coupling of transmission line and resonator and 
no significant change over the measured range in $B$ can be seen.

Both $f_\mathrm{r}(B)$ and $Q_\mathrm{i}(B)$ are perfectly symmetric when 
taking into account the previously determined offset of $B_\mathrm{offs}=\SI{8.5}{\milli\tesla}$.

\section{Two Junction Model}
\label{ap:twojj}
For the derivation of the two junction transmon Hamiltonian, we start with Kirchhoff's current law:
\begin{eqnarray}
C  \ddot{\phi} = - I_\mathrm{c,2} \sin \left( \arctan\frac{\sin\frac{ 2\pi\phi}{\Phi_0}}{r+\cos \frac{2\pi\phi}{\Phi_0}} \right),\nonumber
\end{eqnarray}
where we use the current-phase-relation \fref{eq:curphase} derived in the main part. Without loss of generality, $r = I_{\mathrm{c},2}/ I_{\mathrm{c},1} \geq 1 $ was chosen there.

The Lagrangian for this dynamics is then given by:
\begin{eqnarray}
\mathcal{L} = \frac{C}{2} \dot{\phi}^2 +  \frac{I_\mathrm{c,1}\Phi_0}{2\pi}   \sqrt{r^2+2r\cos\left(  \frac{ 2\pi\phi}{\Phi_0}  \right) +1} .\nonumber
\end{eqnarray}
Introducing the charging energy $E_\mathrm{C} = \frac{\mathrm{e}^2}{2C}$, the Josephson energy $E_\mathrm{J,1} =  \frac{I_\mathrm{c,1}\Phi_0}{2\pi}  $  and the number and phase operators $N$ and $\varphi = \frac{ 2\pi\phi}{\Phi_0}$  with $[N,\varphi] = \mathrm{i}$, we end up with the system Hamiltonian 
\begin{eqnarray}
H = 4 E_\mathrm{C}N^2  - E_\mathrm{J,1}    \sqrt{r^2+2r\cos\varphi+1} .\nonumber
\end{eqnarray}

We now do a Taylor expansion in $\varphi$ to fourth order and neglecting constant terms we get for the approximate Hamiltonian
\begin{eqnarray}
H \approx 4E_\mathrm{C} N^2 +   E_\mathrm{J,1} \left( \frac{r}{2r+2}\varphi^2-\frac{r(r^2-r+1)}{24(r+1)^3}\varphi^4  \right) \nonumber.
\end{eqnarray}
Comparing the harmonic part to a standard quantum harmonic oscillator, we find
\begin{eqnarray}
N &=& i \left(  \frac{\hbar^2}{32}\frac{E_\mathrm{J,1}}{E_\mathrm{C}}\frac{r}{r+1}  \right)^\frac{1}{4} ( a^\dag-a) \nonumber\\
\varphi &=& \left(  \hbar^2 \frac{2E_\mathrm{C}}{E_\mathrm{J,1}} \frac{r+1}{r} \right)^\frac{1}{4}(a^\dag+a)\nonumber.
\end{eqnarray}
Together with the bosonic commutation relation $[a,a^\dag] = 1$ and neglecting all constant terms and terms without pairs of $a$ and $a^\dag$, we get
\begin{eqnarray}
H=\sqrt{8 E_\mathrm{C}   E_\mathrm{J,1}   \frac{r}{r+1}} a^\dag a -
\frac{r^2-r+1}{(r+1)^2}\frac{E_\mathrm{C}}{2} a^\dag a (a^\dag a +1) \nonumber.
\end{eqnarray}

\section{Qubit Transition Frequency for Reduced Superconducting Gap}
\label{ap:gap}
\begin{figure}[tb]
	\includegraphics[width=\columnwidth]{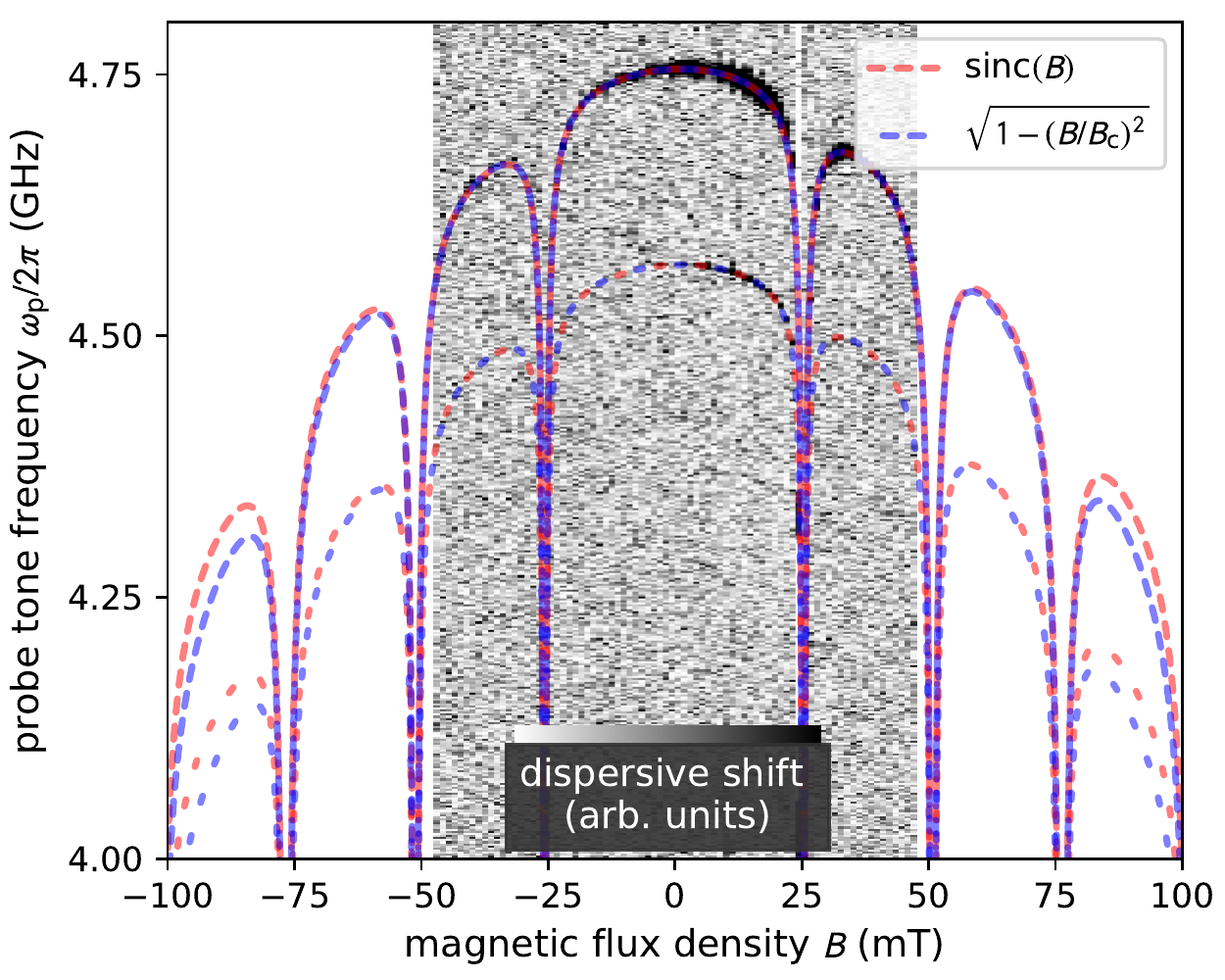}
	\caption{First (narrow-dashed) and second (wide-dashed) qubit transition, calculated using different models of $I_\mathrm{c}$ being reduced by either the magnetic field distribution in the JJ (red) or a decrease of $\Delta(B)$ (blue). In the central part, the measured data are underlaid. It is obvious that with these data, the two effects can not be discriminated.
		Much stronger fields are however not possible with our sample due to the already mentioned reduction of the resonator's quality factor and therefore reduced signal to noise ratio.
	}
	\label{fig:qubit_delta}
\end{figure}
In the main part of the paper, we assume that the overall envelope of $\omega_{01}(B)$ comes from magnetic interference in the main qubit JJ.
However, the overall envelope can also be explained by a reduction of the superconducting gap.
Taking the Ambegaokar-Baratoff relation \cite{Ambegaokar1963}
\begin{eqnarray}
I_c R_n = \frac{\pi}{2e}\Delta(T) \tanh\frac{\Delta(T)}{2 k_B T}\nonumber
\end{eqnarray}
we see that the influence of the $\tanh$ term is negligible for our values of $T$ and $\Delta$, resulting in $I_\mathrm{c} \propto \Delta$. 
Together with $\Delta(B) = \Delta_0 \sqrt{1-(B/B_\mathrm{c})^2}$, where $B_\mathrm{c}$ is the critical field, we get a relation for $I_{\mathrm{c},1}(B)$ and can calculate the qubit transitions. 
To reproduce the same transitions as in the main part, we use $B_\mathrm{c} = \SI{168}{\milli\tesla}$; see the blue line in \fref{fig:qubit_delta}. 
In this limit, $\mathrm{JJ_1}$ is assumed to be point-like, i.e.\@ the periodic Fraunhofer-like reduction of $I_\mathrm{c}$ is neglected.
Within the magnetic field range accessible by our measurements, no deviation from the periodic interference vortex model can be found, and the two effects can not be distinguished with our data.
In reality, both effects coexist at the same time and reduce the critical current.

\section{Modeling Boundaries for $\Gamma_1$}
\label{ap:gamma}
\begin{figure}[t]
	\includegraphics[width=\columnwidth]{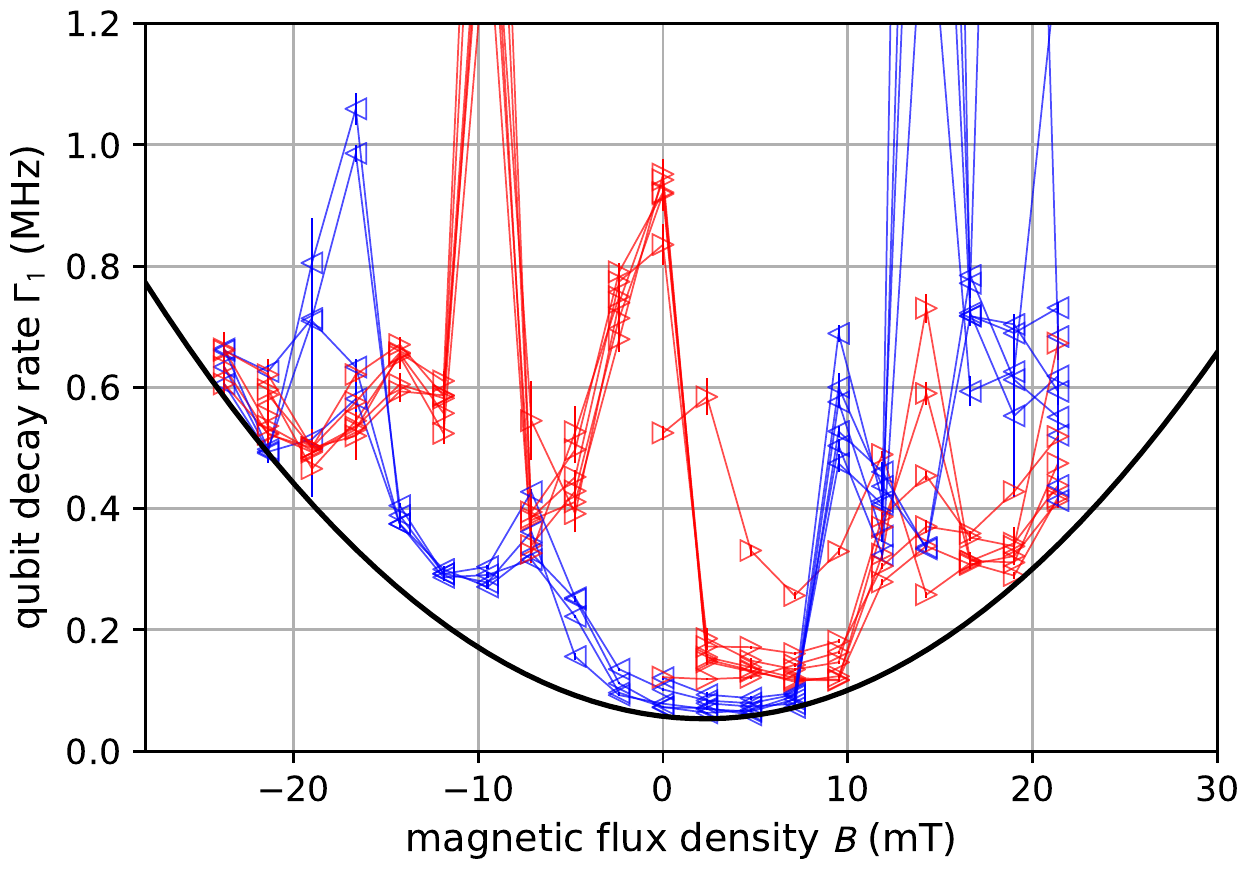}
	\caption{Decay rates of the qubit for the up (red) and down sweeps (blue).
		The data with $\Gamma_1 = 1/T_1$ are the same as used for \fref{fig:coherence_quad} in the main part.
		Shown in black is the parabolic lower limitation of the decay rate as described in the text.
		The lines connect sequentially taken data points and are a guide for the eye.
	}
	\label{fig:qubit_gamma}
\end{figure}
In the main part, the qubit losses are modeled by 
\begin{eqnarray}
\frac{1}{T_1} = \Gamma_1 = \Gamma_\mathrm{hyst} + \Gamma_\mathrm{non\mhyphen hyst} + \Gamma_\mathrm{const},\nonumber
\end{eqnarray}
where $\Gamma_\mathrm{const}$ is assumed to be independent from $B$ and we take the quasiparticle losses $\Gamma_\mathrm{QP} \approx B^2$ as the main contribution to the non-hysteretic $\Gamma_\mathrm{non\mhyphen hyst} $. 
\fref{fig:qubit_gamma} shows our measured data for $\Gamma_1$ together with a lower limitation modeled by $\Gamma = \Gamma_\mathrm{const} + C (B-B_\mathrm{offs})^2$ \cite{Wang2014,Kwon2018,Xi2013}.
We fit the envelope of our data and get $\Gamma_\mathrm{const} = \SI{53.4}{\kilo\hertz}$, $C = \SI{0.785}{\kilo\hertz\per\milli\tesla\squared}$ and $B_\mathrm{offs} = \SI{2.25}{\milli\tesla}$. The remaining hysteretic losses are assumed to come from the entering and movement of flux vortices.

Although the parabola shown in \fref{fig:qubit_gamma}  is a proper envelope for the measured data, we do not want to make any claim that this is a proof for our chosen partitioning of the loss mechanisms. In fact, the different loss mechanisms are not distinguishable by our measurement technique and our partitioning only represents the most obvious loss channels and their dependence on magnetic fields.

\section{Sample and Setup}
\label{ap:setup}
\begin{figure}[t]
	\includegraphics[width=\columnwidth]{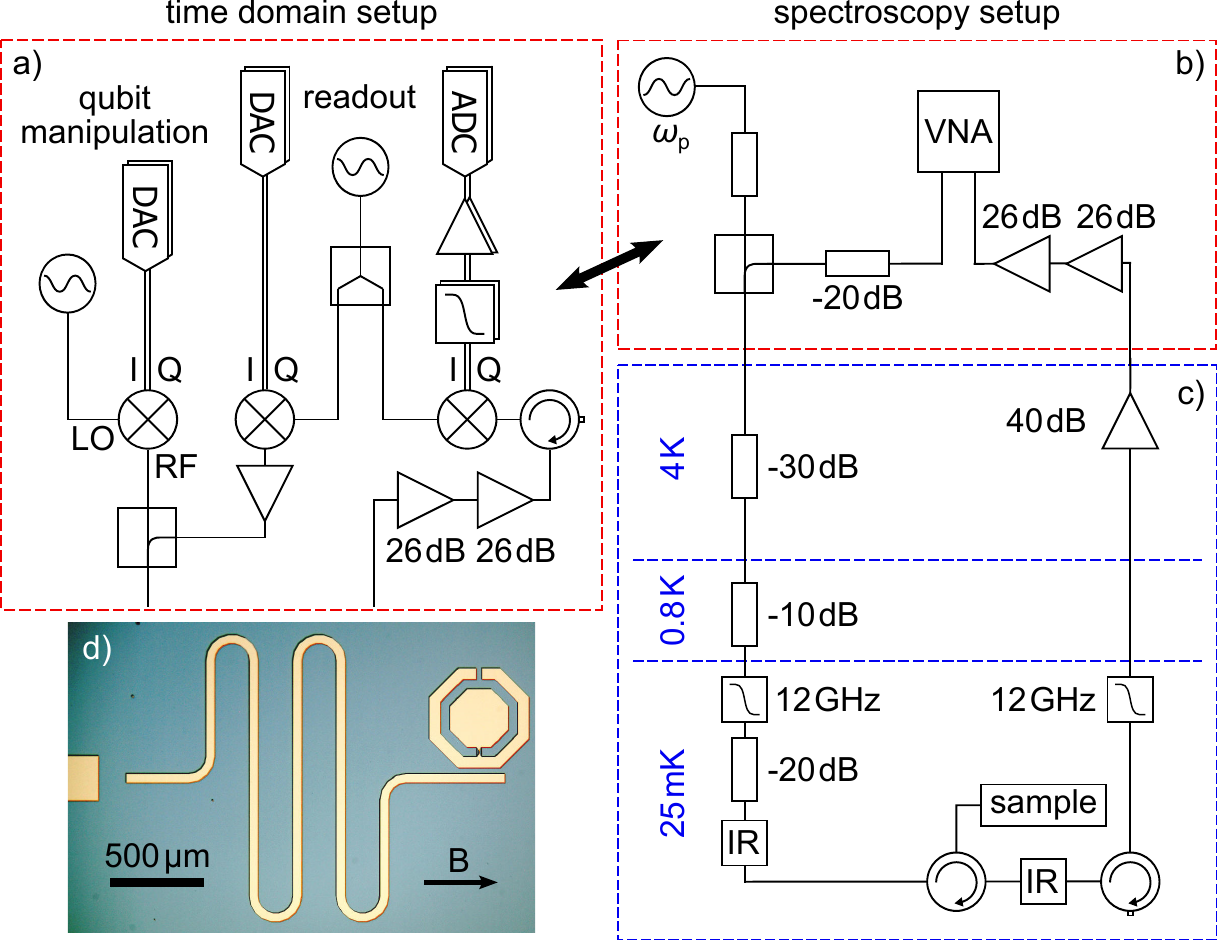}
	\caption{The measurement setup used for data acquisition.
		Either the time domain setup (a) or the spectroscopic setup (b) are connected to the cryostat (c). 
		Both setups feature a readout tone to probe the resonator and a manipulation tone to excite the qubit. The coil around the sample and its leads are not shown here for simplicity.
		(d) Micrograph of the transmon sample including the transmission line (left), resonator (center) and transmon (right). The magnetic field is applied in parallel to the plane of the chip.
	}
	\label{fig:sample}
\end{figure}

For the measurements, we used two different measurement setups: The spectroscopic setup in \fref{fig:sample}b) is used for fast measurements with continuous wave signals, providing a reliable amplitude measurement. This setup was used for measuring the qubit frequency in \fref{fig:two_jj} in the main part and for the additional resonator measurements in \fref{fig:res}.
As the other measurements require pulsed microwave sequences, a home-built time domain setup was used, cf.\@ \fref{fig:sample}a). With the IQ mixers as nonlinear components, this setup does not provide a linear amplitude relation and is therefore not suitable for the calculation of quality factors.

In the cryogenic setup, we attenuate the signal on different stages for thermalization and use microwave cables with low thermal conductance to reduce the heat input to the cold stages, giving a total attenuation of about $\SI{-70}{\deci\bel}$. The reflected signal is amplified by a cryogenic low-noise HEMT amplifier. The sample is shielded from high-frequency noise, infrared radiation and noise from the HEMT by highpass filters, infrared filters, and circulators.

\bibliography{bib}
\bibliographystyle{apsrev4-1}

\end{document}